\documentclass[fleqn,usenatbib]{mnras}

\usepackage{newtxtext,newtxmath}

\usepackage[T1]{fontenc}

\DeclareRobustCommand{\VAN}[3]{#2}
\let\VANthebibliography\thebibliography
\def\thebibliography{\DeclareRobustCommand{\VAN}[3]{##3}\VANthebibliography}

\usepackage{graphicx}	
\usepackage{amsmath}	

\title[Nuclear discs: structure and scaling relations]{Nuclear discs in disc galaxies:  structural properties and scaling relations}

\author[Gadotti \& de S\'a-Freitas]{
Dimitri A. Gadotti$^{1}$\thanks{E-mail: dimitri.a.gadotti@durham.ac.uk}\,and
Camila de S\'a-Freitas$^{2}$\thanks{E-mail: camila.desafreitas@eso.org}
\\
$^{1}$Centre for Extragalactic Astronomy, Department of Physics, Durham University, South Road, Durham DH1 3LE, UK\\
$^{2}$European Southern Observatory, Alonso de C\'ordova 3107, Vitacura, Regi\'on Metropolitana, Chile
}

\date{Accepted XXX. Received YYY; in original form ZZZ}

\pubyear{2024}

\begin{document}
\label{firstpage}
\pagerange{\pageref{firstpage}--\pageref{lastpage}}
\maketitle

\begin{abstract}
The presence of nuclear discs in barred disc galaxies has been demonstrated in studies on stellar structures and kinematics. It is thus imperative to establish their fundamental properties and scaling relations, which can help understanding their connection to other central stellar structures, including nuclear star clusters. In this Letter, we use results from the structural analysis and star formation histories of a sample of galaxies with nuclear discs to show the distributions of fundamental parameters and scaling relations. Our analysis shows that the nuclear disc mass-size relation, and the relation between the nuclear disc mass and the stellar mass of its host galaxy, are -- at face value -- different from the corresponding relations for nuclear star clusters, but marginally compatible given the uncertainties. This sets constraints to scenarios in which their formation is connected. We also find that the nuclear disc in the Milky Way is on the shorter end of the distribution of sizes set by nuclear discs in other galaxies. This is the first analysis of this kind for kinematically confirmed nuclear discs, and further understanding of the properties of the nuclear disc in the Milky Way and other galaxies is necessary to corroborate these results.
\end{abstract}




\section{Introduction}

Structures resembling discs with radii of about several hundred parsecs have long been found in external disc galaxies, particularly with the advent of high resolution imaging data, such as that provided by the Hubble Space Telescope \citep[e.g.,][]{vandenBosch1994,vandenBosch1998,Pizzella2002,Morelli2004}. Often, these structures can be confused with bulges, and were thus included in the family of the so-called `pseudo-bulges' to indicate that their properties are different from those of the classical picture of a galaxy bulge \citep[see, e.g.,][for reviews]{KorKen04,FisDro16}. While some of those early studies attempted at understanding the dynamics of such structures, only more recently, with integral field spectroscopy data, it was possible to ascertain that such structures are indeed rapidly rotating and not simply a continuation of the main galaxy disc towards the very centre \citep[][]{Gadotti2020,Erwin2021}. Altogether, these results show that nuclear discs are ubiquitous. While some nuclear discs are found in elliptical galaxies -- which may constitute a distinct population from nuclear discs in disc galaxies -- in disc galaxies, nuclear discs represent a new paradigm in the context of the central regions of galaxies, as a counterpart to kinematically hot spheroids (i.e., `classical bulges'), and are thought to form from bar-driven gas inflow \citep[][see also \citealt{Schultheis2025} for a review]{ShlFraBeg89,Ath92b,Michel-Dansac2004,Woz07,ColDebErw14,SorBinMag15,SeoKimKwa19}.

It is natural to consider that the gas accumulated at the centre can lead to the formation not only of a nuclear disc but also of other stellar structures, in particular nuclear star clusters, a scenario put forward to explain observations of the nuclear disc and nuclear star cluster in our own Milky Way \citep{Nogueras-Lara2023}. Despite much progress recently in the study of nuclear star clusters \citep[e.g.,][see \citealt{Neumayer2020} for a review]{Fahrion2021}, an exploration of such scenario has only just began.

In this Letter, we leverage on state-of-the-art integral field spectroscopy and on robust, multi-component decompositions of galaxies -- for which the presence of a nuclear disc has been rigorously demonstrated -- to determine fundamental properties and scaling relations of nuclear discs. We show how these compare to known scaling relations of nuclear star clusters and discuss how the Milky Way fits in the picture provided by external galaxies.

\section{Sample and Data Analysis}

The bulk of the results presented here are based on a sample that is essentially the TIMER sample of nearby galaxies, for which there are available both optical integral field spectroscopy data (from MUSE at the Very Large Telescope) and $3.6\mu{\rm m}$ imaging data (from IRAC at the Spitzer Space Telescope). The sample selection is described in detail in \citet{GadSanFal19}. In essence, these galaxies have stellar masses above $10^{10}{\rm M}_\odot$, are at distances below 40\,Mpc, and have inclination angles less than about $60^\circ$. Importantly, all these galaxies are classified -- in the visual morphological classification of \citet{ButSheAth15} -- as being barred {\em and} as having some sort of nuclear structure, such as a nuclear ring, nuclear bar or nuclear spiral arms. In fact, and this is crucial for the present study, the MUSE data has shown that virtually all TIMER galaxies host rapidly rotating central structures (with typical radii of several hundred parsecs), i.e., nuclear discs \citep{Gadotti2020}. It is thus important to stress that these are {\em kinematically confirmed} nuclear discs, and not only structures that visually or photometrically resemble nuclear discs.

After establishing the presence of nuclear discs in these galaxies, to derive their scaling relations it is necessary to estimate their structural parameters (such as mass and size) and here we do this through 2D image decompositions, which account for the contributions of the bar and main disc in these galaxies. The images employed in the decompositions are those from the Spitzer Survey of Stellar Structure in Galaxies \citep[S$^4$G;][]{shereghin10}. All galaxies in this study are part of S$^4$G, which allows for a homogenous analysis, and the depth and wavelength range of this imaging data are key aspects to enhance the accuracy of the decompositions: the depth of the data ensures high signal-to-noise that is particularly beneficial to derive the structural parameters of the main galaxy disc, while the mid-infrared wavelength range avoids effects from dust absorption (particularly relevant in the region where the nuclear disc dominates) and is more representative of the bulk of the stellar populations. However, it should be pointed out that emission from heated dust in this wavelength range can be important in regions of elevated star formation \citep[e.g.,][]{QueMeiSch15}. In addition, with a Point Spread Function (PSF) Full Width at Half Maximum (FWMH) of $\approx1.7''$, the spatial resolution of these images is relatively poor, but this is not an issue in this study, given the angular sizes of the galaxies. In fact, the poorest physical spatial resolution in the sample corresponds to NGC\,7140 ($\approx300\,{\rm pc}$) which is enough to resolve its nuclear disc with a diameter of $\approx700\,{\rm pc}$.

The decompositions were performed with {\sc imfit} \citep{Erwin2015} and are described in detail in a companion paper (Gadotti 2025, MNRAS, subm.; G25). In contrast to common practice, these decompositions were performed employing the Differential Evolution (DE) algorithm to find the best fit to the galaxies, as opposed to the Levenberg-Marquardt algorithm. While the latter requires subjective initial guesses for the the fitted parameters and are more prone to find localised, non-optimal solutions, the DE fits are more likely to find solutions closer to or at the maximum optimisation of the model. In addition, the DE fits do not require initial guesses of the fitted parameters, but rather lower and upper limits, within which the algorithm searches for many candidate solutions, which minimises subjective choices in the fits.

All galaxies were fitted with a model consisting of a S\'ersic bar\footnote{G25 shows that the S\'ersic indices of the bar and photometric ``bulge'' are largely independent and often significantly different, as differences in size and geometrical properties aid in their separation during the fit.}, an exponential main disc and a S\'ersic photometric ``bulge'' \citep{Ser68}. The latter thus encompasses all possible structures present in the central region of the galaxy. However, in \citet{Gadotti2020} and \citet{Bittner2020}, we show that in virtually all galaxies in the TIMER sample, the central region is dominated by the nuclear disc, and only in IC\,1438 and NGC\,1291 there is suggesting evidence that a small classical bulge (i.e., a small kinematically hot spheroid) is embedded within the nuclear disc. Even in this case, the candidate classical bulge is significantly more compact than the nuclear disc \citep[see][]{deLSanMen19}, although it can still clearly influence the optimal model for the S\'ersic photometric ``bulge'' in this galaxy.  From the 21 original TIMER galaxies, however, we have excluded from the photometric analysis NGC\,1365, NGC\,5236 and NGC\,5248 (because their image decompositions are particularly difficult due to various factors), NGC\,7552 (whose IRAC data seems to suffer from saturation at the very centre) and NGC\,6902 (whose central structure is kinematically peculiar).

Therefore, the sample employed in most of the results shown below consists of 16 galaxies for which the structural properties of the main components that make up the galaxies were carefully estimated. In the context of this study, the relevant structural parameters are the S\'ersic index $n$ of the nuclear disc, its effective (or half-light) radius, and the luminosity fraction of the galaxy contained in the nuclear disc. From the latter we can obtain the stellar mass in the nuclear disc using the galaxy stellar mass as in Table\,1 of \citet{GadSanFal19} and multiplying it by the nuclear disc luminosity fraction, including a correction for mass-to-light ratio variations across the galaxy using the procedure formulated in \citet{QueMeiSch15}. A corresponding calculation is carried out to derive the bar stellar mass discussed below. Another parameter discussed below is the kinematic radius of the nuclear disc, $r_k$, which is a characteristic radius measured in \citet{Gadotti2020} and is the defined as the radius along the major axis of the nuclear disc where $v/\sigma$ peaks ($v$ and $\sigma$ being, respectively, the stellar line-of-sight velocity and velocity dispersion).

To expand on the photometric analysis, when noted, we also include in this study the stellar masses and sizes of nuclear discs derived with the MUSE spectroscopy data in \citet[][dSF+25]{deSa-Freitas2025}. The stellar masses are computed from an analysis of the star formation histories (SFH) in different regions in the sample galaxies. The method starts from the principle that, in the region of the galaxy where the nuclear disc resides, other stellar structures co-exist, which are collectively referred to as the main underlying disc. Furthermore, it is assumed that the SFH of the nuclear disc can be isolated by subtracting the SFH of the main underlying disc from the directly observed SFH in the central region of the galaxy, which includes therein both the nuclear disc and the main underlying disc. For that, the SFH of the main underlying disc is determined just outside the nuclear disc and assuming a model for the main disc of the galaxy \citep[for details, see][]{deSa-Freitas2023,deSa-Freitas2023b}. The nuclear disc sizes are estimated from visual inspection of radial profiles of mean stellar age, velocity dispersion and specific star formation rate \citep[see][]{deSa-Freitas2025}. This spectroscopic analysis is done with 18 galaxies from the TIMER sample, plus two galaxies with archival MUSE data that host the smallest nuclear discs confirmed to date. NGC\,1291 is excluded, since its nuclear disc fills the MUSE field of view, making the analysis of its SFH overly challenging. There are thus 15 TIMER galaxies that are included in both the photometric and spectroscopic analyses presented here.

Throughout this Letter, to convert angular sizes to physical sizes, we used the galaxy distances as tabulated in Table\,1 of \citet{GadSanFal19}, and a Hubble constant of ${\rm H_0}=67.8\,\rm{km}\,\rm{s}^{-1}\,\rm{Mpc}^{-1}$ and $\Omega_{\rm m}=0.308$ in a universe with flat topology \citep{AdeAghArn15}.

\section{Structural Properties and Scaling Relations}

\begin{figure*}
\includegraphics[trim=0.2cm 0.2cm 0.2cm 0.2cm,clip=true,width=2\columnwidth]{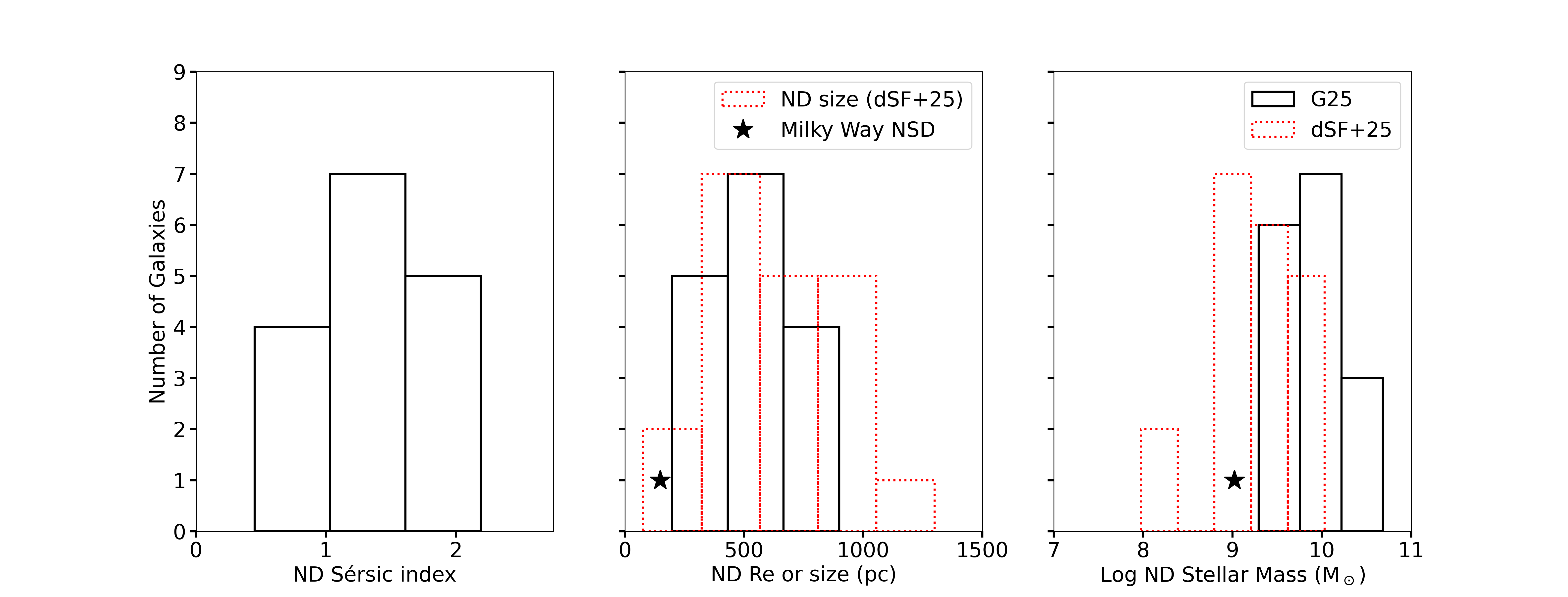}
\caption{Distribution of nuclear disc structural parameters obtained through the photometric decompositions (G25) and spectroscopic analysis \citep[the latter from][dSF+25]{deSa-Freitas2025}. {\it Left:} S\'ersic index; {\it middle:} effective radius (in black) and size (red dotted lines), and {\it right:} stellar mass. The star on the latter two panels shows the position of the Milky Way nuclear disc from the results presented in \citet{Sormani2022}.}
\label{fig:Dists}
\end{figure*}

In Fig.\,\ref{fig:Dists}, we show the distributions of the nuclear disc S\'ersic index, effective radius $R_e$ and spectroscopic size, and stellar mass (both photometric and spectroscopic). The typical S\'ersic index of a nuclear disc is thus seen to be in the range between $\approx0.5$ and $\approx2$ and peaking at $\approx1.5$, confirming that they are well described with close to exponential radial density profiles (i.e., $n=1$), just as main galaxy discs. A comparison of the values obtained here and those obtained by \citet{SalLauLai15} for the same galaxies, and also using the same S$^4$G imaging data, shows a very good agreement (see companion paper, G25). The distribution of $R_e$ shows that it spans the range $\approx200-900\,{\rm pc}$ with a peak around $\approx500\,{\rm pc}$. The distribution of the sizes derived spectroscopically is wider, although with a similar peak. It is important to note that, while $R_e$ and $r_k$ are characteristic radii, the spectroscopic sizes are defined as the outer limits of the nuclear discs. The nuclear disc stellar masses measured photometrically and spectroscopically clearly do not agree, but are correlated, with the latter being systematically lower by about 0.5 dex (see discussion below).
The nuclear disc in the Milky Way is thought to have a scale length of $h\approx89\,{\rm pc}$ and stellar mass $\approx10.5\times10^8{\rm M}_\odot$ \citep{Sormani2022}. These values were derived via dynamical modelling. The position of the Milky Way nuclear disc is shown with a star in Fig.\,\ref{fig:Dists}, after converting scale length to effective radius assuming an exponential profile and the relation provided in \citet[][$R_e=1.678\times h$, when $n=1$]{GraDri05}. The Milky Way nuclear disc is clearly in the short size end of the distribution shown here. Its mass is also low when masses are measured photometrically, but falls well within the distribution of spectroscopic masses. Nevertheless, we point out that the parameters concerning the Milky Way were derived employing different modelling techniques.

\begin{figure}
\includegraphics[trim=0.2cm 0.2cm 0.2cm 0.2cm,clip=true,width=\columnwidth]{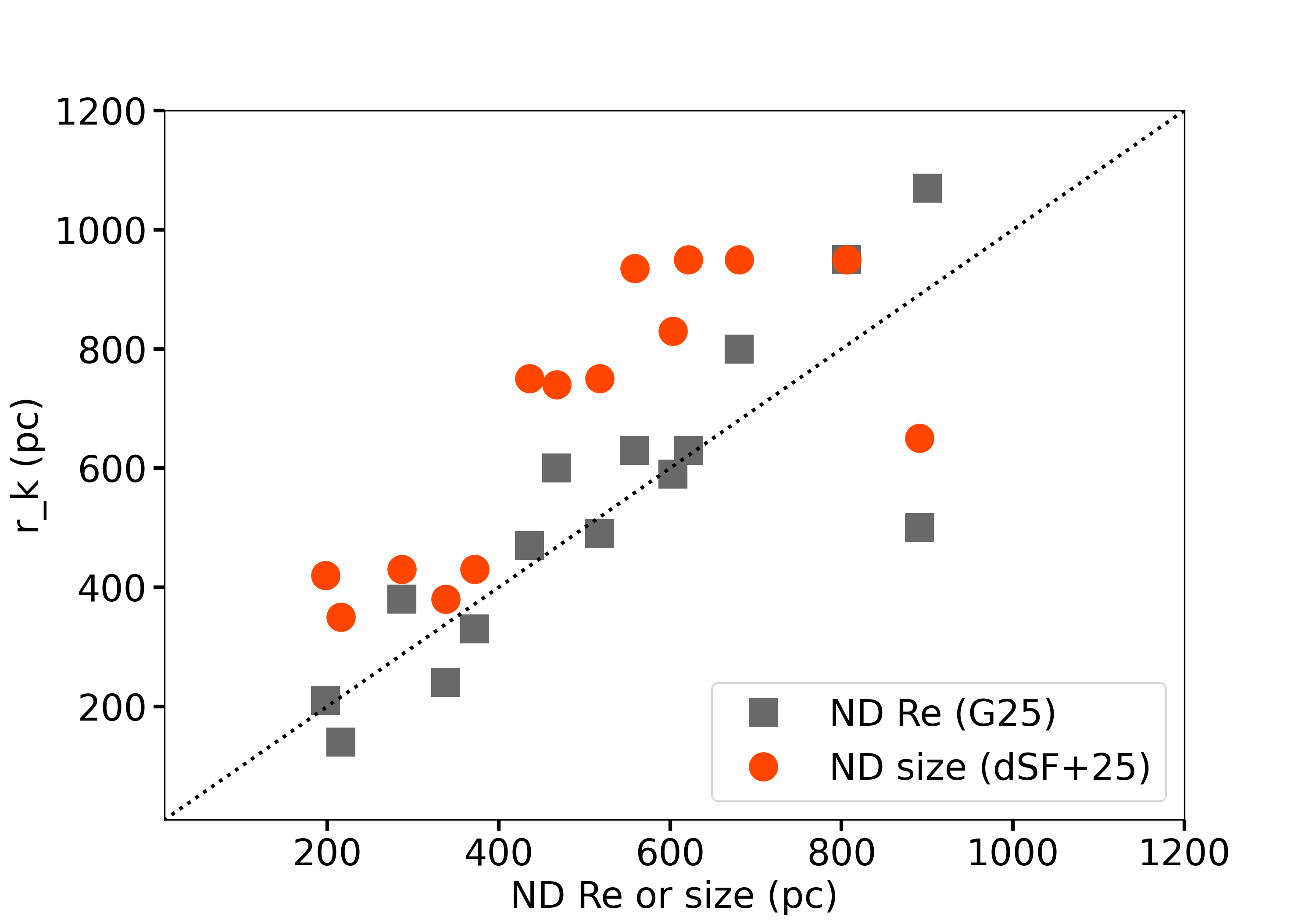}
\caption{Nuclear disc kinematic radius ($r_k$) as a function of the effective radius ($R_e$) of the photometric `bulge' component in the decompositions (squares). Circles denote the outer edges of the nuclear discs derived in \citet[][dSF+25]{deSa-Freitas2025}. The dotted line is not a fit to the data, but rather depicts a one-to-one relation, and is an excellent representation of the actual $r_k-R_e$ relation. This is corroborating evidence that the central photometric component (or photometric `bulge') is physically a rapidly rotating nuclear disc.}
\label{fig:sizes}
\end{figure}

It is interesting to explore how the kinematically defined radius $r_k$ relates with $R_e$. This was done in \citet{Gadotti2020}, but using decompositions available then in the literature. In that study it was found that the two parameters correlate well, and that the ratio $r_k/R_e$ was clustered at $\approx0.8$. We revisit this relation here with the more homogenous decompositions presently available and show the result in Fig.\,\ref{fig:sizes}. The figure confirms the strong correlation between the two radii, but, strikingly, now one sees that a one-to-one relation (depicted by the dotted line in the figure) represents very well the correlation. In other words, $r_k/R_e\approx1$, which is strong evidence that photometric exponential ``bulges'' are nuclear discs, and that $r_k$ does not necessarily marks the outer edge of the nuclear disc (but is a characteristic radius), unless the latter has a strongly truncated profile. In fact, the nuclear disc sizes derived spectroscopically (i.e., their outer edges) also correlate very well with $r_k$, but are systematically larger, as expected. The increase in the average value of $r_k/R_e$ from $\approx0.8$ to $\approx1$ is likely due to the new decompositions in G25 using an oversampled PSF, which, as shown in G25, leads to more accurate estimates of the S\'ersic index and $R_e$. Some of the literature $R_e$ values used previously were obtained without employing an oversampled PSF, and thus are likely overestimated.

\begin{figure*}
\includegraphics[trim=0.2cm 0.2cm 0.2cm 0.2cm,clip=true,width=2\columnwidth]{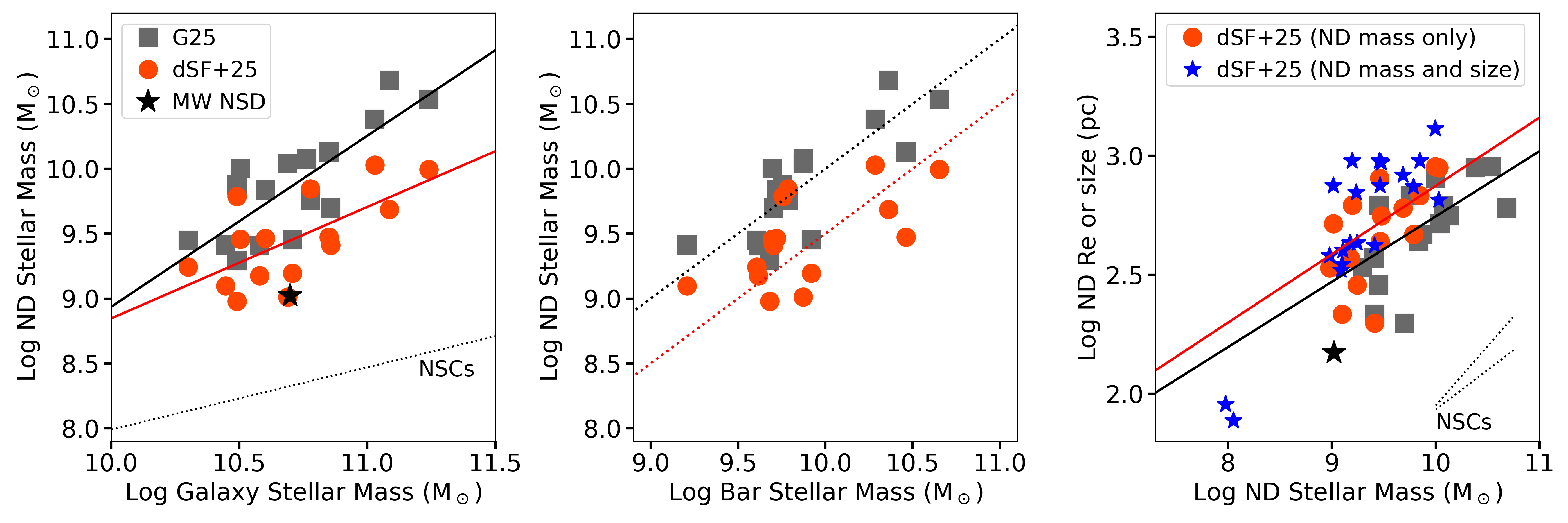}
\caption{Scaling relations of nuclear discs. {\it Left:} relation between the stellar mass in the nuclear disc and the total stellar mass of the host galaxy. {\it Middle:} relation between the nuclear disc stellar mass and the bar stellar mass. {\it Right:} relation between the nuclear disc effective radius and its stellar mass (i.e., the nuclear disc mass-size relation). Filled black squares correspond to nuclear disc stellar masses measured photometrically (G25), whereas the filled red circles correspond to the spectroscopic measurements of the nuclear disc stellar mass (dSF+25; the latter include only the 15 galaxies for which photometric decompositions are also available). The solid lines are fits to the relations in the left and right panels, as described in the text. The black dotted line in the middle panel depicts a one-to-one relation and is an excellent representation of the actual relation, whereas the red dotted line is a one-to-one relation shifted by 0.5 dex. The star on the left and right panels shows the position of the Milky Way nuclear disc from the results presented in \citet{Sormani2022} and \citet{Bland-Hawthorn2016}. The Milky Way is not included in the fits. In the right panel, the blue stars show the mass-size relation of nuclear discs when using {\em both} mass and size as derived spectroscopically for the full sample in \citet[][dSF+25]{deSa-Freitas2025}. Finally, the dotted lines in the left and right panels indicate the relations followed by nuclear star clusters (NSCs). In the left panel, this corresponds to Eq. 1 in \citet{Neumayer2020}, but with the NSC masses multiplied by 10 to aid visualisation. In the right panel, the dotted segments indicate the slopes of 0.33 and 0.5, within which lies the NSC mass-size relation discussed in \citet{Neumayer2020}.}
\label{fig:SRels}
\end{figure*}

Having the the sizes and masses of the nuclear discs, we can now derive some important scaling relations; these are shown in Fig.\,\ref{fig:SRels}, using total stellar masses for the galaxies as in Table 1 of \citet{GadSanFal19}. For the Milky Way, the mass and size of the nuclear disc are taken from \citet{Sormani2022}, with the mass of the Galaxy as quoted in \citet{Bland-Hawthorn2016}. We can see (left panel) that the mass of the nuclear disc is correlated with the mass of the host galaxy, and this is true for both the photometric and spectroscopic measurements of the nuclear disc stellar mass, despite their significant offset. The black solid line is a power law fit to the data using the photometric masses, which yields:

\begin{equation}
\log M_{\rm ND} = \left(1.32\pm0.25\right) \log \left(\frac{M_{\rm Gal}}{10^9\,{\rm M}_\odot}\right) + \left(7.62\pm0.44\right),
\label{eq:mgal}
\end{equation}

\noindent where $M_{\rm ND}$ and $M_{\rm Gal}$ are the nuclear disc and host galaxy stellar masses, respectively, and the uncertainties are derived via bootstrapping. On the other hand, the corresponding fit that arises when using the spectroscopic masses (red solid line) is:

\begin{equation}
\log M_{\rm ND} = \left(0.86\pm0.25\right) \log \left(\frac{M_{\rm Gal}}{10^9\,{\rm M}_\odot}\right) + \left(7.99\pm0.44\right).
\label{eq:mgalCdSF}
\end{equation}

In the middle panel of Fig.\,\ref{fig:SRels}, we show the relation between the nuclear disc stellar mass and the bar stellar mass.
The black dotted line is not a fit to the data but simply a depiction of a one-to-one relation. Stunningly, the latter is an excellent fit to the data, i.e., when using the results from the photometric decompositions only, the nuclear disc and the bar have roughly the same stellar mass! This is typically about 10 per cent of the host galaxy stellar mass (cf. left panel in the figure). This is {\em not} true when the spectroscopic measurements of the nuclear disc stellar mass are used: the correlation between the bar and the nuclear disc stellar masses remains, but the nuclear disc stellar masses are now systematically lower by about 0.5 dex (red dotted line). It is interesting to note that the {\em sizes} of nuclear discs and bars in the TIMER sample are also strongly correlated (see Fig.~8 in \citealt{Gadotti2020} and Fig.~1 in dSF+25; see also \citealt{ComKnaBec10} for earlier work with a different sample, where a correlation is also found, albeit with larger scatter).

The right panel of Fig.\,\ref{fig:SRels} shows the nuclear disc mass-size relation. The black solid line is a power law fit to the data using the photometric masses, yielding:

\begin{equation}
R_e \propto M_{\rm ND}^{(0.274\pm0.003)},
\label{eq:masssize}
\end{equation}

\noindent with the uncertainty again derived via bootstrapping. The corresponding fit that arises when using the spectroscopic masses (red solid line) is:

\begin{equation}
R_e \propto M_{\rm ND}^{(0.287\pm0.003)}.
\label{eq:masssizeCdSF}
\end{equation}

\noindent Note that none of the fits presented here includes the Milky Way. Also shown is the mass-size relation of nuclear discs when using {\em both} mass and size as derived spectroscopically for the full sample in \citet{deSa-Freitas2025}.

In the context of a possible connection in the formation of nuclear discs and nuclear star clusters, it is instructive to compare the above relations to the corresponding relations for nuclear star clusters as derived in \citet[][see their Figs.\,7 and 12 and Eqs.\,1 and 2]{Neumayer2020}. These authors find a slope for the relation between the mass of the nuclear star cluster and the mass of the host galaxy of 0.48 (with an uncertainty found via bootstrapping of 0.04), which is almost a factor 3 lower than the slope we find in Eq.\,\ref{eq:mgal} (1.32), but closer to the one we find in Eq.\,\ref{eq:mgalCdSF} (0.86), albeit still different at the 1.5$\sigma$ level. However, the shallower slope for nuclear star clusters is derived when considering clusters in both early- and late-type galaxies, and while both early- and late-type galaxies seem to follow the same cluster-galaxy mass relation, there are hints that the slope is steeper for clusters in late-type galaxies only. When restricting the fit to clusters with more accurate mass measurements -- which are the most massive clusters and primarily in late-type galaxies -- \citet{Neumayer2020} find a steeper slope, namely $0.92\pm0.18$. This is of course intermediate between the values we find in Eqs.\,\ref{eq:mgal} and \ref{eq:mgalCdSF}.

\citet{Neumayer2020} also derived the slope of the mass-size relation for nuclear star clusters, finding that it lies between 0.33 and 0.5, which is again, on average, different and significantly higher than the values we find in Eqs.\,\ref{eq:masssize} (0.27; with photometric masses) and \ref{eq:masssizeCdSF} (0.29; with spectroscopic masses). In addition, it is interesting that \citeauthor{Neumayer2020} found that below a certain mass, nuclear star clusters have a fixed radius independent of the mass, which is not seen here for nuclear discs, although this could result from the fact that the nuclear discs studied here do not reach masses low enough to show this feature.

\citet{Georgiev2016} also studied these relations for nuclear star clusters and found slopes in line with the results of \citet{Neumayer2020}, and thus different from the corresponding slopes for nuclear discs. Specifically, they found that the slope in Eqs.\,\ref{eq:mgal}/\ref{eq:mgalCdSF} for nuclear star clusters is $0.356^{+0.056}_{-0.057}$ for late-type galaxies and $0.326^{+0.055}_{-0.051}$ for early-type galaxies. And for Eqs.\,\ref{eq:masssize}/\ref{eq:masssizeCdSF} for nuclear star clusters they found a slope of $0.321^{+0.047}_{-0.038}$ for late-type galaxies and $0.347^{+0.024}_{-0.024}$ for early-type galaxies. In the latter case, the estimated uncertainties put the result in line to what is found here, at least when the nuclear disc stellar masses are derived using the SFHs. Altogether, this comparison between the scaling relations of nuclear discs and nuclear star clusters reveals that the relations are -- at face value -- distinct. However, statistically, they are still compatible, considering the uncertainties involved.



\section{Implications and Concluding Remarks}

The results presented above corroborate previous results, indicating not only that nuclear discs show exponential profiles (or close to exponential; $n\sim1-2$), just as main discs, but also that such structures are retrieved as exponential photometric `bulges' in image decompositions. Astonishingly, the new decompositions discussed here show an excellent one-to-one correlation between the corresponding kinematic and photometric radii.

And for the first time for kinematically confirmed nuclear discs, here we also present some scaling relations for nuclear discs in disc galaxies. However, an accurate derivation of some of these scaling relations is hampered by the difficulties in estimating the stellar mass of nuclear discs. In fact, the results above show that the masses derived from photometric decompositions are systematically larger by about 0.5 dex than the values derived via an analysis of the central SFHs. While it is reasonable to expect that the true values lie between these two measurements, it is reassuring that the measurements are correlated. In fact, there is an excellent one-to-one correlation between the masses of the nuclear discs and their progenitor bars when the former come from photometry, and the correlation remains when using the SFHs to derive the stellar masses of the nuclear discs, albeit with a significant offset. In this case, the nuclear disc stellar mass is typically about 30 per cent of the mass of the bar. The latter values are more in line with our theoretical understanding of the dynamics of barred galaxies, as too strong central concentrations of mass may destroy bars \citep[see, e.g.,][]{Hasan1993,AthLamDeh05}.

In addition, the mass--size relation and the relation between the nuclear disc and its host galaxy masses follow power laws (with the mass-size relation possibly a broken power law or a more curved relation). Both power laws differ from the corresponding relations concerning nuclear star clusters, although their mass-size relations are marginally compatible given the uncertainties, and the relations are closer when using nuclear disc stellar masses derived spectroscopically. That does not necessarily mean that these stellar systems are built in fundamentally different processes, but successful models of the formation of both nuclear discs and nuclear star clusters must reproduce such observed differences -- if such differences are confirmed with future work involving larger samples. Since bars funnel gas towards the central region of their host galaxies, it is natural to envisage a scenario in which the gas inflow can promote the building of a nuclear star cluster at the same time as it builds the nuclear disc, at least in the initial stages of the nuclear disc build-up. Further in time, the inflow of gas is halted at the edges of the nuclear disc (often resulting in a star-forming nuclear ring) and the growing of the nuclear star cluster may continue on a different path. Such a scenario would possibly result in diverse scaling relations, as observed. It should also be noted that nuclear discs grow in mass as they age \citep[as shown in][]{deSa-Freitas2025}, and so they may move in the relations shown.

Finally, our results highlight that the Milky Way nuclear disc is in the short end in the distribution of sizes of confirmed nuclear discs in disc galaxies. Evidently, the obscured view we have of the Milky Way's nuclear disc makes such measurements challenging. Progress in this regard to either confirm or revise the current measurements in the Milky Way and external galaxies would be most welcome, particularly on the stellar masses of nuclear discs, such that the results presented here can be put on firmer grounds.

\section*{Acknowledgements}

We are grateful to the anonymous referee for the very helpful comments. We are thankful to Mathias Schultheis, Mattia Sormani, Francesca Fragkoudi and Miguel Querejeta for fruitful discussions. DAG is supported by STFC grants ST/T000244/1 and ST/X001075/1. This work used the DiRAC@Durham facility managed by the Institute for Computational Cosmology on behalf of the STFC DiRAC HPC Facility (www.dirac.ac.uk). The equipment was funded by BEIS capital funding via STFC capital grants ST/K00042X/1, ST/P002293/1, ST/R002371/1 and ST/S002502/1, Durham University and STFC operations grant ST/R000832/1. DiRAC is part of the National e-Infrastructure.

\section*{Data Availability}

All original $3.6\mu$m S$^4$G images used in this study are available at \href{https://irsa.ipac.caltech.edu/data/SPITZER/S4G/}{https://irsa.ipac.caltech.edu/data/SPITZER/S4G/}.



\bibliographystyle{mnras}
\bibliography{/Users/dgadotti/work/papers/gadotti_refs} 

\bsp	
\label{lastpage}
\end{document}